\documentclass[sigconf]{acmart}
\usepackage{multirow}
\usepackage{subfigure}
\usepackage{enumerate}
\usepackage{enumitem}
\usepackage{stfloats}
\usepackage{hyperref}
\usepackage{breakurl}
\usepackage{colortbl}
\usepackage{diagbox}
\usepackage{balance}
\usepackage{titlesec}
\usepackage{verbatimbox}
\usepackage{wrapfig}
\usepackage{stfloats}
\usepackage{tablefootnote}
\usepackage[linesnumbered,ruled,vlined]{algorithm2e}

\usepackage{cleveref}
\usepackage[most]{tcolorbox}
\usepackage{lipsum}
\usepackage{tikz}

\usepackage{graphicx}

\usepackage[export]{adjustbox}

\crefname{section}{§}{§§}

\definecolor{best}{RGB}{255, 159, 69}
\definecolor{gray}{RGB}{221, 221, 221}

\AtBeginDocument{%
  }

\definecolor{cell}{RGB}{191, 234, 245}
\definecolor{highlight}{RGB}{220, 53, 53}
\definecolor{darkblue}{RGB}{20, 66, 114}

\def\up{\color{blue} }
\def\down{\color{red}}

\definecolor{block1}{rgb}{0.2, 0.2, 0.2}
\definecolor{content}{RGB}{245, 245, 245}

\def\model{WSFE}
\def\longmodel{Wasserstein Sub-graph Feature Encoder}

\SetKwComment{Comment}{$\triangleright$\ }{}

\makeatletter
\newcommand\notsotiny{\@setfontsize\notsotiny\@vipt\@viipt}
\makeatother

\newcommand\mth{\small}

\newenvironment{sequation}{\begin{equation}\setlength\abovedisplayskip{2pt}\setlength\belowdisplayskip{2pt}}{\end{equation}}

\def\emb{\boldsymbol} 

\DeclareMathOperator\argmin{argmin}

\setcopyright{acmcopyright}
\copyrightyear{2023}
\acmYear{2023}
\setcopyright{acmlicensed}\acmConference[SIGIR '23]{Proceedings of the 46th International ACM SIGIR Conference on Research and Development in Information Retrieval}{July 23--27, 2023}{Taipei, Taiwan}
\acmBooktitle{Proceedings of the 46th International ACM SIGIR Conference on Research and Development in Information Retrieval (SIGIR '23), July 23--27, 2023, Taipei, Taiwan}
\acmPrice{15.00}
\acmDOI{10.1145/3539618.3592089}
\acmISBN{978-1-4503-9408-6/23/07}

\begin{document}
\fancyhead{}


\title{\model: Wasserstein Sub-graph Feature Encoder for Effective User Segmentation in Collaborative Filtering}

\author{Yankai Chen}
\affiliation{%
  \country{The Chinese University of Hong Kong}
}
\email{ykchen@cse.cuhk.edu.hk}

\author{Yifei Zhang}
\affiliation{%
  \country{The Chinese University of Hong Kong}
}
\email{yfzhang@cse.cuhk.edu.hk}

\author{Menglin Yang}
\affiliation{%
  \country{The Chinese University of Hong Kong}
}
\email{mlyang@cse.cuhk.edu.hk}

\author{Zixing Song}
\affiliation{%
  \country{The Chinese University of Hong Kong}
}
\email{zxsong@cse.cuhk.edu.hk}

\author{Chen Ma}
\affiliation{%
  \country{City University of Hong Kong}
}
\email{chenma@cityu.edu.hk}

\author{Irwin King}
\affiliation{%
  \country{The Chinese University of Hong Kong}
}
\email{king@cse.cuhk.edu.hk}


\begin{abstract}
Maximizing the user-item engagement based on vectorized embeddings is a standard procedure of recent recommender models.
Despite the superior performance for item recommendations, these methods however implicitly deprioritize the modeling of \textit{user-wise similarity} in the embedding space;
consequently, identifying similar users is underperforming, and additional processing schemes are usually required otherwise.
To avoid thorough model re-training, we propose {\model}, a model-agnostic and training-free representation encoder, to be flexibly employed on the fly for effective user segmentation.
Underpinned by the optimal transport theory, the encoded representations from \model~ present a matched user-wise similarity/distance measurement between the realistic and embedding space.
We incorporate \model~ into six state-of-the-art recommender models and conduct extensive experiments on six real-world datasets.
The empirical analyses well demonstrate the superiority and generality of \model~to fuel multiple downstream tasks with diverse underlying targets in recommendation. 

\end{abstract}

\maketitle

\section{Introduction}
Collaborative filtering (CF), as one effective strategy to perform personalized modeling and prediction, has been widely deployed for recommendation.
One prevalent learning paradigm of CF models~\cite{lightgcn,pinsage,ling2012online,chen2022learning,yang2022hicf,wu2020semi} is to parameterize users and items as vectorized embeddings and learn to reconstruct users' historical interactions.
As such, the learned embeddings are convenient to interpret target users' diverse preferences and predict their future behaviors.

In addition to reflecting preferences on items, another desirable property of learned user embeddings is to explicitly capture the \textit{user-wise similarity}; this provides an intuitive recognition of similar user interests and affinities, which lays the foundation and is particularly beneficial for \textit{user-centric} analyses and applications such as group recommendation and advertising~\cite{xin2023SIRecs,xin2023Optimization}.
However, this property is usually deprioritized and neglected by recent models~\cite{lightgcn,ngcf,sgl,buir,yang2022hrcf,luodiscrete}.
To address the unsatisfactory performance in similar user identification, thorough model re-training may thus be required. 
To tackle this issue, we are motivated to encode high-quality embeddings in collaborative filtering, such that they can efficiently and seamlessly serve the task of user segmentation.

\begin{figure}[t]
\begin{minipage}{0.5\textwidth}
\includegraphics[width=3.3in]{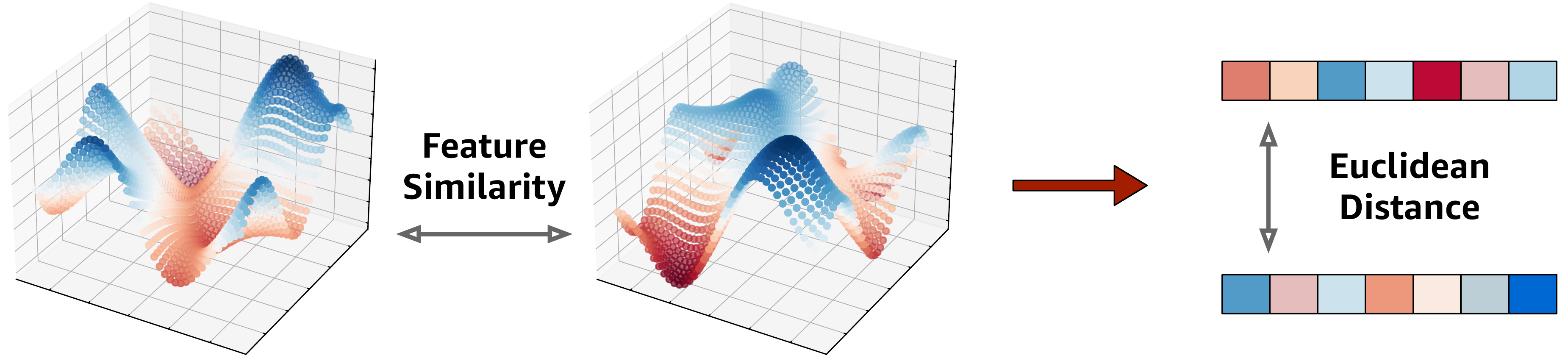}
\end{minipage} 
\caption{Illustration of \model~ in encoding the similarity of empirical feature distribution (left) with the corresponding embedding distance (right).}
\label{fig:intro}
\end{figure}

In this work, we propose \underline{W}asserstein \underline{S}ub-graph \underline{F}eature \underline{E}ncoder (\model), to explicitly model the user behaviors in the form of user-item interaction graph, and measure the user-wise similarity by exploiting their high-order sub-graph patterns.
We notice that users with similar interaction behaviors naturally share overlapping sub-graph patterns.
Based on this observation, one straightforward solution would be to exhaustively calculate similarities for all the nodes in underlying sub-graphs; this however may be intractable in practice mainly because of the exponential node scale in graph exploration.
On the contrary, our proposed \model~captures user similarity by directly encoding their sub-graph latent features, enabling it model-agnostic and flexible for a variety of graph-based recommender models.
Specifically, as shown in Figure~\ref{fig:intro}, we assume the user preference follows an unknown high-dimensional probability distribution; this unique preference distribution is \textit{empirically observed and represented by the latent features that are well-learned in the item recommendation task}.
Then \model~explicitly captures the distribution distances with {Wasserstein metrics} from the optimal transport theory~\cite{rabin2011wasserstein,villani2009optimal,pswe,WEGL}.
Consequently, the encoded user representations can effectively reflect their realistic item-interaction similarity, producing a matched Euclidean distance measurement for ease of user segmentation in the embedding space.

To summarize, our contributions are highlighted as follows:
\begin{itemize}[leftmargin=*]
\item To the best of our knowledge, we are the first to focus on improving the embedding quality for effective user segmentation in collaborative filtering, while not jeopardizing the model evaluation for item recommendation.

\item We propose \model~for effective representation encoding via capturing the feature similarity of high-order user-item interaction graph patterns.
\model~ is adaptive for any graph-based models and training-free; thus it can be invoked on the fly as long as the backbone models are well-trained.

\item We conduct extensive experiments by fusing \model~into six state-of-the-art models on six real-world datasets.
Not only do we present its performance superiority in empirical evaluation, but we also provide technical discussion for future investigation.
\end{itemize}




\section{\model~Methodology}
\label{sec:method}

\subsection{Preliminaries}
\label{sec:pre}

\textbf{Graph-based Collaborative Filtering.}
In view of user-item interaction graphs, the general idea of graph-based approaches is to capture CF signals in high-hop neighbors.
In this work, we study the \textit{Graph Convolutional Networks (GCNs)} to learn node representations by smoothing the latent features via topology~\cite{kipf2016semi,chen2023bipartite,zixing1,zixing2}. 
It iteratively propagates neighborhood information to the target node, e.g., user $u$, which can be abstracted:
\begin{sequation}
\emb{v}_{ngh\rightarrow u}^{(l)} = Prop\Big(\{\emb{v}_i^{(l-1)}: i \in \mathcal{N}(u)\}\Big),
\end{sequation}%
where {\mth $\emb{v}_{ngh\rightarrow u}^{(l)}$} is the representation after $l$ layers of propagation from interacted items in $u$'s neighboring set {\mth $\mathcal{N}(u)$}.
With the propagated information, node embeddings are iteratively updated by aggregating features of the center and neighbor nodes~\cite{graphsage,zhang2022knowledge,he2023dynamically}.

\noindent\textbf{Optimal Transport and Wasserstein Metrics.}
Optimal transport is the general problem of moving one distribution of mass, e.g., {\mth $P$}, to another, e.g., {\mth $Q$}, as efficiently as possible.
The derived minimum {\mth $L_2$} cost can be referred as their distribution distance:
\begin{sequation}
\label{eq:OT_definition}
W_2(P, Q) = \Big({\inf}_{f\in \mathcal{F}(P,Q)} \int \|\emb{x}-f(\emb{x})\|^2 d P(\emb{x})\Big)^{\frac{1}{2}},
\end{sequation}%
where the infimum is over all transport plans in {\mth $\mathcal{F}$} between {\mth $P$} and {\mth $Q$}.
For \textit{one-dimensional} distributions, there is a closed-form solution to compute such \textit{optimal transport map} {\mth$f^*$} as \underline{\mth $f^*(x):=F_{P}^{-1}\big(F_{Q}(x)\big)$}; $F$ is the cumulative distribution function (CDF) associated with $P$.

For the \textit{high-dimensional} case, the metric of \textit{sliced-Wasserstein distance}~\cite{rabin2011wasserstein,bonneel2015sliced,deshpande2019max} is formally defined as follows:
\begin{sequation}
\label{eq:sw2}
SW_2\left(P, Q\right)=\Big(\int_{\mathbb{S}^{d-1}} W_2^2\big(P^{\emb\theta}, Q^{\emb\theta}\big) d \mth{\emb\theta}\Big)^{\frac{1}{2}},
\end{sequation}%
where {\mth $P^{\emb\theta}$} is projected by function {\mth $g^{\emb\theta}$: $\mathbb{R}^d$ $\rightarrow$ $\mathbb{R}$} as {\mth $P^{\emb\theta}$ := $g^{\emb\theta}(P)$} and {\mth $g^{\emb\theta}(\mth{x}) = {\mth {\emb\theta}^\mathsf{T}x} $}. ${\mth \emb\theta}$ is a unit vector in $\mathbb{R}^d$ and {\mth $\mathbb{S}^{d-1}$} is the unit $d$-dimensional hypersphere.
Due to holding \textit{positive-definiteness}, \textit{symmetry}, and \textit{triangle inequality}~\cite{kolouri2016sliced,kolouri2019generalized,pswe,zhang2020discrete,xinfish}, we employ it as the distance measurement for high-dimensional subgraph feature distributions.

\subsection{Sub-graph Feature Encoding}

\subsubsection{\textbf{Formulating Sub-graph Feature Distributions.}}
As illustrated in Figure~\ref{fig:subgraph}(A), if two users are considered to be similar in terms of historical preferences, they should share similar behaviors with overlapping interaction graph patterns.
Based on this intuition, consider that each user's preference follows an unknown, independent, and $d$-dimensional probability measure, e.g., $P_u$.
We assume that the interaction pattern of $u$ observed so far is sampled from the underlying distribution $P_u$.
Thus the \textit{empirical (discrete) distribution} {\mth $\widehat{P}_u$} with its empirical CDF can be formulated as:
\begin{sequation}
\label{eq:ecdf}
F_{\widehat{P}_u}(\emb{x}) = \frac{1}{L+1} {\sum}_{l=0}^{L} \delta(\emb{x} - \emb{v}_{ngh\rightarrow u}^{(l)}).
\end{sequation}%
Notice that we initialize {\mth $\emb{v}_{ngh\rightarrow u}^{(0)}$ $:=$ $\emb{v}_{u}^{(0)}$}.
Here {\mth $\delta(\cdot)$} returns 1 if the input is zero and 0 otherwise (discretizing from continuous case {\mth $\int$$\delta(x)dx$ $=$ $1$}).
Without loss of generality, these empirical distributions are representative, i.e., {\mth $\widehat{P}_u$ $\approx$ $P_u$}; thus we would refer {\mth $P_u$} to {\mth $\widehat{P}_u$} hereafter to avoid notation abuse.

\subsubsection{\textbf{Implementing $f^*$.}}
We first set a $d$-dimensional \textit{reference distribution} {\mth $P_o$} that functions as the ``origin'' in the embedding space to measure the distance toward any inputs.
{\mth $P_o$} is associated with random feature embeddings, e.g., {\mth $\emb{v}_{o}^{(l)}$}, as: {\mth $F_{{P}_o}(\emb{x}) = \frac{1}{L+1} {\sum}_{n=0}^{L} \delta(\emb{x} - \emb{v}_{o}^{(l)}) $}.
To implement the optimal transport map {\mth $f^*$} for such discrete and $d$-dimensional case, we have the following procedure.

\begin{figure}[t]
\hspace{-0.1in}
\begin{minipage}{0.5\textwidth}
\includegraphics[width=3.3in]{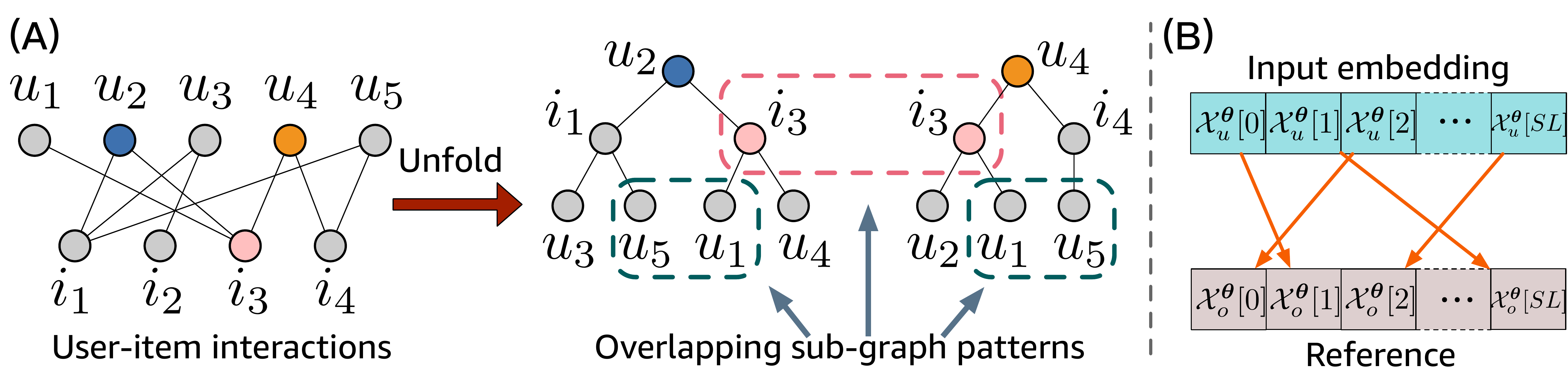}
\end{minipage} 
\vspace{-0.1in}
\caption{(A) Illustration of the similar sub-graph patterns. (B) Mapping processing between the input and reference.}
\label{fig:subgraph}
\end{figure}
We first conduct distribution slicing to {\mth $P_o$} and {\mth $P_u$} by projection function {\mth $g^{\emb\theta}$}.
For each pair of distribution slices {\mth $P_u^{\emb\theta}$} and {\mth $P_o^{\emb\theta}$}, let {\mth $\mathcal{X}_u^{\emb\theta}$} collect their projected sub-graph features as {\mth $\mathcal{X}_u^{\emb\theta}$ = $\{\emb\theta^{\mathsf{T}} \emb{v}_{ngh\rightarrow u}^{(l)} \}_{l=0}^L$} (so does for {\mth $\mathcal{X}_o^{\emb\theta}$ = $\{\emb\theta^{\mathsf{T}} \emb{v}_{o}^{(l)} \}_{l=0}^L$}).
Then the corresponding optimal transport map {\mth $f^*(x):=F_{P_u^{\emb\theta}}^{-1}\big(F_{P_o^{\emb\theta}}(x)\big)$} can be quantitatively intepreted:
\begin{sequation}
{f}^*(x|\mathcal{X}_u^{\emb\theta}) = \argmin_{x' \in \mathcal{X}_u^{\emb\theta}} \big(F_{P_u^{\emb\theta}}(x') = r\big) \text{, where } r = F_{P_o^{\emb\theta}}(x).
\end{sequation}%
Furthermore, let {\mth $\tau(x' |\mathcal{X}_u^{\emb\theta})$} denote the ranking of each input {\mth $x'$} in the \textit{ascending sorting} of {\mth $\mathcal{X}_u^{\emb\theta}$}. We can replace the term {\mth $F_{P_u^{\emb\theta}}$} and have:
\begin{sequation}
\label{eq:f}
f^*(x | \mathcal{X}_u^{\emb\theta})=\operatorname{argmin}_{x' \in \mathcal{X}_u^{\emb\theta}}\big(\tau(x' | \mathcal{X}_u^{\emb\theta})=\tau(x | \mathcal{X}_o^{\emb\theta})\big).
\end{sequation}%

As shown in Figure~\ref{fig:subgraph}(B), Eqn.(\ref{eq:f}) essentially permutes different layers of sub-graph embeddings of {\mth $\mathcal{X}_u^{\emb\theta}$} in encoding, such that the distance to the reference of {\mth $\mathcal{X}_o^{\emb\theta}$} can be subsequently captured and embedded.
Please notice that the distance is in the $L_2$-norm form as shown in Eqn.(\ref{eq:OT_definition}), \textit{a.k.a.} the Euclidean distance, which is favorable to scenarios for recalling vectorized objects that requires a reasonable distance measurement in the embedding space.

\subsubsection{\textbf{Implementing \model.}}
For each pair of distribution slices, based on the algorithmic implementation of Eqn.(\ref{eq:f}), we proceed to encode their representations as follows:
\begin{sequation}
\label{eq:layerwise}
\emb{E}_u^{\emb\theta} := \frac{1}{L+1} \Big{\|}_{l=0}^{L} f^*\Big({\emb{\theta}}^{\mathsf{T}}\emb{v}_o^{(l)}|\mathcal{X}_u^{\emb\theta}\Big) - \emb{E}_o^{\emb\theta} \text{  and  } \emb{E}_o^{\emb\theta} := \frac{1}{L+1} \Big{\|}_{l=0}^{L} {\emb{\theta}}^{\mathsf{T}}\emb{v}_o^{(l)},
\end{sequation}%
where $||$ denotes the concatenation operation.
According to the theory in Eqn.(\ref{eq:sw2}), the next step is to draw infinite projections for distance integral, which, however, may be computationally expensive and infeasible in practice.
In this work, we implement it with Monte-Carlo approximation with $S$ times of uniform sampling from {\mth $\mathbb{S}^{d-1}$}.
Consequently, this leads to a cumulative sliced-Wasserstein distance (i.e., approximating Eqn.(\ref{eq:sw2})) between reference {\mth $P_o$} and the original input feature distribution {\mth $P_u$} as:
\begin{sequation}
\label{eq:sum_sw2}
S W_2(P_o, P_u) \approx\Big(\frac{1}{S} {\sum}_{s=1}^S W_2^2\big(P_o^{\theta_s}, P_u^{\theta_s}\big)\Big)^{\frac{1}{2}}.
\end{sequation}%

Regularized by the distance cumulation in Eqn.(\ref{eq:sum_sw2}), our \textit{\longmodel~(\model)} is finally defined:
\begin{sequation}
\label{eq:wsfe}
\emb{E}_u := \frac{1}{S} \Big{\|}_{s=1}^{S} \emb{E}_u^{\emb\theta_s} \text{ \quad and \quad } \emb{E}_o := \frac{1}{S} \Big{\|}_{s=1}^{S} \emb{E}_o^{\emb\theta_s},
\end{sequation}%
where {\mth $\emb{E}_u$, $\emb{E}_o$ $\in$ $\mathbb{R}^{S(L+1)}$}.
Notice that in practice, the number of graph convolutions {\mth $L$ $\leq$ 4}~\cite{lightgcn,kipf2016semi,graphsage} is a common setting mainly to avoid the \textit{over-smoothing problem}~\cite{li2019deepgcns}.
Moreover, our empirical observations in \cref{sec:change_S} reveal that setting {\mth $S=64$} already achieves satisfactory model performances with an acceptable computational cost.

\subsection{Theoretical Analysis}

\begin{wrapfigure}[8]{r}{0.22\textwidth}
\vspace{-0.05in}
\includegraphics[width=1.5in]{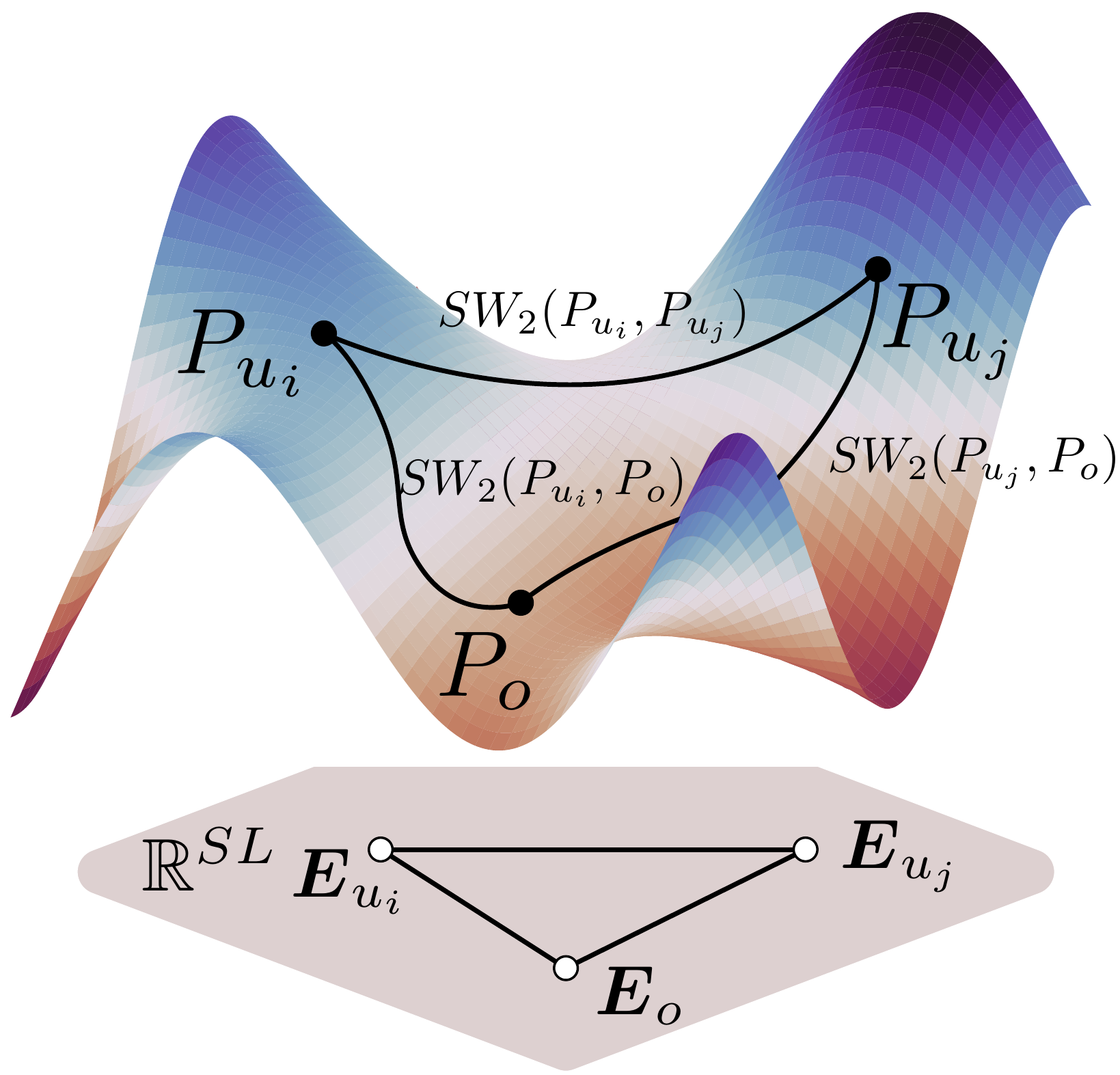}
\vspace{-0.15in}
\caption{Wasserstein to Euclidean geodesics.}
\label{fig:geo}
\end{wrapfigure}

One major expectation of encoded representations is that they can reflect the similarity/distance of their sub-graph feature distributions.
We illustrate this in Figure~\ref{fig:geo} with the theorem as follows:

\noindent\textsc{\textbf{Theorem 1.}}
\textit{For any input sub-graph features of users $u_i$ and $u_j$ with distributions {\mth $P_{u_i}$} and {\mth $P_{u_j}$}, their encoded representations hold:}
\\
\textcircled{\raisebox{-0.7pt}{1}}  {\mth $\|\emb{E}_{u_i}\|_2$ $\approx$ $SW_2(P_{u_i}, P_0)$}. \\
\textcircled{\raisebox{-0.7pt}{2}} {\mth $\|\emb{E}_{u_i} - \emb{E}_{u_j}\|_2$ $\approx$ $SW_2(P_{u_i}, P_{u_j})$}.

\ 

\noindent\textsc{Proof.}
The proof is twofold.
For property \textcircled{\raisebox{-0.7pt}{2}}, we have:\\
\begin{equation}
\begin{aligned}
 & \quad \|\emb{E}_{u_i} - \emb{E}_{u_j}\|_2 \\
& = \Big\| \big{\|}_{s=1}^{S} (\emb{E}_{u_i}^{\emb{\theta}_s} - \emb{E}_{u_j}^{\emb{\theta}_s}) \Big\|_2 \\ 
&=  \Big\| {\sum}_{s=1}^{S} (\emb{E}_{u_i}^{\emb{\theta}_s} - \emb{E}_{u_j}^{\emb{\theta}_s})^2 \Big\|_2\\ 
&=  \Big\| \frac{1}{S} {{\sum}}_{s=1}^{S} \frac{1}{L+1} {\sum}_{l=0}^{L} \left({f^{*}}(\emb{\theta}_s^{\mathsf{T}}\emb{v}^{(l)}_{o}|\mathcal{X}_{{u_i}}^{\emb{\theta}_s}) - {f^{*}}(\emb{\theta}_s^{\mathsf{T}}\emb{v}^{(l)}_{o}|\mathcal{X}_{u_j}^{\emb{\theta}_s}) \right)^2 \Big\|_2\\
& \approx  \Big\| \int_{\mathbb{S}^{d-1}} \int_{\mathbb{R}} \Big(F^{-1}_{P_{u_i}^{\emb\theta}}\big(F_{P_o^{\emb\theta}}(t)\big) - F^{-1}_{P_{u_j}^{\emb\theta}}\big(F_{P_o^{\emb\theta}}(t)\big) \Big)^2  dP_o^{\emb\theta}(t) d{\emb\theta} \Big\|_2\\
&=  \Big\| \int_{\mathbb{S}^{d-1}} \int_0^1\big(F_{P_{u_i}^{\emb\theta}}^{-1}(r)-F_{P_{u_j}^{\emb\theta}}^{-1}(r)\big)^2 dr d{\emb\theta}\Big\|_2.
\end{aligned}
\end{equation}%
Let $F^{-1}_{P_{u_i}^{\emb\theta}}(r)$ = $x$, meaning that $r$ = $F_{P_{u_i}^{\emb\theta}}$. We have:
\begin{equation}
\begin{aligned}
 \quad \|\emb{E}_{u_i} - \emb{E}_{u_j}\|_2 &=  \Big\| \int_{\mathbb{S}^{d-1}} \int_0^1\big(x - F_{P_{u_j}^{\emb\theta}}^{-1}(F_{P_{u_i}^{\emb\theta}}(x))\big)^2 dP_{u_j}^\theta(x) d{\emb\theta}\Big\|_2 \\
& =  \Big\|\int_{\mathbb{S}^{d-1}}  W_2^2(P_{u_j}^{\emb\theta}, P_{u_i}^{\emb\theta}) d{\emb\theta} \Big\|_2  = SW_2(P_{u_j}, P_{u_i}). 
\end{aligned}
\end{equation}%
With symmetry, we have {\mth $SW_2(P_{u_i}, P_{u_j})$ = $SW_2(P_{u_j}, P_{u_i})$}.
\noindent Then for the reference $P_o$, its encoded representation is straightforward to have {\mth $\emb{E}_o$ = $\emb{0}$}.
Thus we complete the proof as follows:
\begin{sequation}
\|\emb{E}_{u_i}\|_2 = \|\emb{E}_{u_i} - \emb{E}_o\|_2 \approx SW_2(P_{u_i}, P_o),
\end{sequation}%
$\hfill\blacksquare$

\noindent\textbf{Complexity Analysis.}
\model~is \textit{training-free} that can be utilized on the fly right after the backbone model is trained.
Thus, the complexity of \model~ is {\mth $O\big(S(L+1)M {d}\log{d}\big)$}, where the cost {\mth $O(d\log{d})$} is for implementing {\mth $\tau(\cdot)$}.
Fortunately, it is linear to the input data size, indicating that the encoding can be done at the input scale.

\section{Experimental Results}
\label{sec:exp}
We evaluate \model~with the aim of answering the following research questions:
\textbf{RQ1.} How does \model~ boost the user segmentation performance of state-of-the-art recommender models?
\textbf{RQ2.} How do different model settings affect \model~ performance?

\subsection{Experimental Setups}

\textbf{Datasets.}
We collect six widely-evaluated public datasets (including their original training/test data splits) from: MovieLens~\cite{movie,chen2022modeling,chen2022attentive,zhang2022knowledge}, Gowalla~\cite{cho2011friendship,gowalla}, Pinterest~\cite{geng2015learning,pst}, Yelp~\cite{yelp}, Kindle~\cite{kindle,yu2022self}, and Amazon-Book~\cite{book,lightgcn}.
Dataset statistics are reported in Table~\ref{tab:datasets}.

\begin{table}[t]
\centering
\footnotesize
\caption{The statistics of datasets.}
\vspace{-0.15in}
\label{tab:datasets}
\setlength{\tabcolsep}{0.7mm}{
\begin{tabular}{c | c | c | c | c | c | c}
\toprule 
             & {\scriptsize MovieLens}  & {\scriptsize Gowalla}   & {\scriptsize Pinterest}  &  {\scriptsize Yelp2018} & {\scriptsize Kindle} &{\scriptsize AMZ-Book} \\
\midrule[0.1pt]
\midrule[0.1pt]
    {\footnotesize \#Users }  & {6,040}   & {29,858}   & {55,186}   & {31,668}   &{115,652}  &{52,643}\\ 
    {\footnotesize \#Items }  & {3,952}   & {40,981}   & {9,916}    & {38,048}    &{98,729}  &{91,599}\\
  	{\footnotesize \#Avg. Interactions } & {165.60} & {34.31} & {26.52} & {49.31} &{15.85} & {56.69} \\
  	{\footnotesize \#All Interactions } & {1,000,209} & {1,027,370} & {1,463,556} & {1,561,406}  & {1,833,068} & {2,984,108}  \\
\bottomrule
\end{tabular}}
\end{table}

\textbf{Evaluation Protocol.}
The fundamental property required by user segmentation is \textit{user-wise similarity
measurement}.
Thus, given a query user, we treat this task as ranking towards candidates of similar users, based on the encoded user representations. 
In this work, we sort out similar users based on the number of overlapping items they have interacted with; then we compared these ranking lists with Top-K results inferred from the learning models.
Recall@K and NDCG@K are the evaluation metrics.

\textbf{Baselines.}
To demonstrate the effectiveness of \model, we incorporate it into the following state-of-the-art models.
\begin{enumerate}[leftmargin=*]
\item \textbf{LightGCN}~\cite{lightgcn} is one state-of-the-art GCN-based recommender model with a more concise and powerful structure.


\item \textbf{SGL}~\cite{sgl} is one representative graph-based model with contrastive learning to tackle the data sparsity issue.

\item \textbf{SimGCL}~\cite{simgcl} is the state-of-the-art contrastive-learning-based recommender model that conducts the simplified augmentation directly in the feature space.

\item \textbf{NCL}~\cite{ncl} is one of the state-of-the-art graph-based models with contrastive neighborhood information enrichment.

\item \textbf{BUIR}~\cite{buir} is one state-of-the-art model that bootstraps user and item representations for collaborative filtering.

\item \textbf{DirectAU}~\cite{directau} is the latest model that improves the representation quality from the perspective of alignment and uniformity.

\end{enumerate}

\subsection{Empirical Analyses and Discussions}
\subsubsection{\textbf{Overall Performance (RQ1).}}
\label{sec:overall}

\begin{table*}[t]
\centering
\scriptsize
\caption{Experimental results before and after implementing \model~ into the underlying models (best view in color).}
\vspace{-0.15in}
\label{tab:exp}
\setlength{\tabcolsep}{0.2mm}{
\begin{tabular}{c | c| c c | c c | c c | c c   }
\toprule 
     Dataset    & Model  &   {Recall@5}   &  {NDCG@5}   &  {Recall@20}   &  {NDCG@20}  &  {Recall@50}  &  {NDCG@50}   &  {Recall@100}   &  {NDCG@100}  \\
\midrule[0.1pt]

\multirow{7}{*}{Movie} & {LightGCN}  & {0.10{\tiny$\rightarrow$}0.18 ({\up +80.00\%})}  & {0.36{\tiny$\rightarrow$}0.61 ({\up +69.44\%})}  	& {0.37{\tiny$\rightarrow$}0.69 ({\up +86.49\%})}  & {0.36{\tiny$\rightarrow$}0.65 ({\up +80.56\%})}   & {0.85{\tiny$\rightarrow$}1.68 ({\up +97.65\%})}  & {0.62{\tiny$\rightarrow$}1.20 ({\up +93.55\%})}  & {1.63{\tiny$\rightarrow$}3.04 ({\up +86.50\%})}   & {0.97{\tiny$\rightarrow$}1.83 ({\up +88.66\%})} \\ 


~ & {SGL}  & {0.08{\tiny$\rightarrow$}0.13 ({\up +62.50\%})}  & {0.32{\tiny$\rightarrow$}0.47 ({\up +46.88\%})}  	& {0.24{\tiny$\rightarrow$}0.37 ({\up +54.17\%})}  & {0.25{\tiny$\rightarrow$}0.39 ({\up +56.00\%})}  & {0.42{\tiny$\rightarrow$}0.73 ({\up +73.81\%})}  & {0.36{\tiny$\rightarrow$}0.59 ({\up +63.89\%})}  & {0.60{\tiny$\rightarrow$}1.22 ({\up +103.33\%})}  & {0.47{\tiny$\rightarrow$}0.81 ({\up +72.34\%})}     \\ 

~ & {SimGCL}  & {0.90{\tiny$\rightarrow$}1.02 ({\up +13.33\%})}  & {3.40{\tiny$\rightarrow$}3.04 ({\up +13.82\%})}  	& {2.13{\tiny$\rightarrow$}3.77 ({\up +77.00\%})}  & {2.30{\tiny$\rightarrow$}3.60 ({\up +56.52\%})}  & {3.10{\tiny$\rightarrow$}8.56 ({\up +176.13\%})}  & {2.92{\tiny$\rightarrow$}5.99 ({\up +105.14\%})}  & {4.45{\tiny$\rightarrow$}13.71 ({\up +208.09\%})}  & {3.35{\tiny$\rightarrow$}8.58 ({\up +156.12\%})}     \\ 

~& {NCL}  & {0.25{\tiny$\rightarrow$}0.38 ({\up +52.00\%})}  & {0.84{\tiny$\rightarrow$}1.30 ({\up +54.76\%})}  	& {0.88{\tiny$\rightarrow$}1.47 ({\up +67.05\%})}  & 0.84{{\tiny$\rightarrow$}1.38 ({\up +64.29\%})}  & {2.01{\tiny$\rightarrow$}3.42 ({\up +70.15\%})}  & {1.47{\tiny$\rightarrow$}2.47 ({\up +68.03\%})}  & {3.60{\tiny$\rightarrow$}6.19 ({\up +71.94\%})}  & {2.20{\tiny$\rightarrow$}3.74 ({\up +70.00\%})}     \\ 

~ & {BUIR}  & {0.22{\tiny$\rightarrow$}0.24 ({\up +9.09\%})}  & {0.74{\tiny$\rightarrow$}0.82 ({\up +10.82\%})}  	& {0.83{\tiny$\rightarrow$}0.96 ({\up +15.66\%})}  & {0.79{\tiny$\rightarrow$}0.92 ({\up +16.46\%})}  & {1.81{\tiny$\rightarrow$}2.19 ({\up +20.99\%})}  & {1.34{\tiny$\rightarrow$}1.60 ({\up +19.40\%})}  & {3.21{\tiny$\rightarrow$}3.95 ({\up +23.05\%})}  & {1.98{\tiny$\rightarrow$}2.40 ({\up +21.21\%})}     \\ 

~ & {DirectAU}  & {0.09{\tiny$\rightarrow$}0.10 ({\up +11.11\%})}  & {0.32{\tiny$\rightarrow$}0.36 ({\up +12.50\%})}  	& {0.25{\tiny$\rightarrow$}0.31 ({\up +24.00\%})}  & {0.26{\tiny$\rightarrow$}0.32 ({\up +23.08\%})}  & {0.53{\tiny$\rightarrow$}0.66 ({\up +24.53\%})}  & {0.41{\tiny$\rightarrow$}0.51 ({\up +24.39\%})}  & {0.98{\tiny$\rightarrow$}1.23 ({\up +25.51\%})}  & {0.62{\tiny$\rightarrow$}0.77 ({\up +24.19\%})}     \\ 

\midrule[0.1pt]

\multirow{7}{*}{Gowalla} & {LightGCN}  & {4.41{\tiny$\rightarrow$}4.56 ({\up +3.40\%})}   & {6.59{\tiny$\rightarrow$}6.52 ({\down -1.06\%})}     & {8.75{\tiny$\rightarrow$}9.26 ({\up +5.83\%})}    & {6.64{\tiny$\rightarrow$}6.80 ({\up +2.41\%})}    & {13.30{\tiny$\rightarrow$}14.05 ({\up +5.64\%})}    & {8.46{\tiny$\rightarrow$}8.71 ({\up +2.96\%})}    & {17.78{\tiny$\rightarrow$}18.86 ({\up +6.07\%})}    & {9.94{\tiny$\rightarrow$}10.29 ({\up +3.52\%})}   \\ 


~ & {SGL}  & {4.96{\tiny$\rightarrow$}5.30 ({\up +6.85\%})}  & {7.58{\tiny$\rightarrow$}7.62 ({\up +0.53\%})}  	& {9.85{\tiny$\rightarrow$}10.66 ({\up +8.22\%})}  & {7.53{\tiny$\rightarrow$}7.86 ({\up +4.38\%})}  & {15.13{\tiny$\rightarrow$}16.68 ({\up +10.24\%})}  & {9.66{\tiny$\rightarrow$}10.23 ({\up +5.90\%})}  & {20.70{\tiny$\rightarrow$}22.72 ({\up +9.76\%})}  & {11.49{\tiny$\rightarrow$}12.19 ({\up +6.09\%})}     \\ 

~ & {SimGCL}  & {5.25{\tiny$\rightarrow$}6.94 ({\up +32.19\%})}  & {8.55{\tiny$\rightarrow$}10.32 ({\up +20.70\%})}  	& {10.14{\tiny$\rightarrow$}13.99 ({\up +37.97\%})}  & {8.11{\tiny$\rightarrow$}10.52 ({\up +29.72\%})}  & {14.73{\tiny$\rightarrow$}20.23 ({\up +37.34\%})}  & {9.99{\tiny$\rightarrow$}13.02 ({\up +30.33\%})}  & {17.88{\tiny$\rightarrow$}24.15 ({\up +35.07\%})}  & {11.12{\tiny$\rightarrow$}14.34 ({\up +28.96\%})}     \\

~ & {NCL}  & {4.65{\tiny$\rightarrow$}5.01 ({\up +7.53\%})}  & {8.31{\tiny$\rightarrow$}8.71 ({\up +4.81\%})}  	& {9.41{\tiny$\rightarrow$}10.20 ({\up +8.40\%})}  & {7.81{\tiny$\rightarrow$}8.34 ({\up +6.79\%})}  & {13.78{\tiny$\rightarrow$}15.11 ({\up +9.65\%})}  & {9.71{\tiny$\rightarrow$}10.43 ({\up +7.42\%})}  & {18.16{\tiny$\rightarrow$}20.02 ({\up +10.24\%})}  & {11.22{\tiny$\rightarrow$}12.11 ({\up +7.93\%})}     \\

~ & {BUIR}  & {2.94{\tiny$\rightarrow$}3.07 ({\up +4.42\%})}  & {5.64{\tiny$\rightarrow$}5.74 ({\up +1.77\%})}  	& {6.65{\tiny$\rightarrow$}6.92 ({\up +4.06\%})}  & {5.52{\tiny$\rightarrow$}5.66 ({\up +2.54\%})}  & {10.74{\tiny$\rightarrow$}11.00 ({\up +2.42\%})}  & {7.30{\tiny$\rightarrow$}7.42 ({\up +1.64\%})}  & {15.09{\tiny$\rightarrow$}15.36 ({\up +1.79\%})}  & {8.78{\tiny$\rightarrow$}8.91 ({\up +1.48\%})}     \\ 

~ & {DirectAU}  & {5.03{\tiny$\rightarrow$}5.41 ({\up +7.55\%})}  & {7.99{\tiny$\rightarrow$}8.22 ({\up +2.88\%})}  	& {10.24{\tiny$\rightarrow$}10.98 ({\up +7.23\%})}  & {7.90{\tiny$\rightarrow$}8.31 ({\up +5.19\%})}  & {15.71{\tiny$\rightarrow$}16.94 ({\up +7.83\%})}  & {10.16{\tiny$\rightarrow$}10.74 ({\up +5.71\%})}  & {21.20{\tiny$\rightarrow$}22.86 ({\up +7.83\%})}  & {12.00{\tiny$\rightarrow$}12.73 ({\up +6.08\%})}     \\ 
\midrule[0.1pt]

\multirow{7}{*}{Pinterest} & {LightGCN}  & {2.24{\tiny$\rightarrow$}2.38 ({\up +6.25\%})}  & {4.64{\tiny$\rightarrow$}4.94 ({\up +6.47\%})}  	& {7.24{\tiny$\rightarrow$}7.68 ({\up +6.08\%})}  & {5.51{\tiny$\rightarrow$}5.87 ({\up +6.53\%})}  & {13.67{\tiny$\rightarrow$}14.67 ({\up +7.32\%})}  & {8.37{\tiny$\rightarrow$}8.94 ({\up +6.81\%})}  & {21.11{\tiny$\rightarrow$}22.65 ({\up +7.30\%})}  & {11.04{\tiny$\rightarrow$}11.81 ({\up +6.97\%})}     \\


~ & {SGL}  & {3.93{\tiny$\rightarrow$}3.92 ({\down -0.25\%})}  & {7.58{\tiny$\rightarrow$}7.58 ({0\%})}  	& {11.21{\tiny$\rightarrow$}11.35 ({\up +1.25\%})}  & {8.57{\tiny$\rightarrow$}8.63 ({\up +0.70\%})}  & {19.62{\tiny$\rightarrow$}19.86 ({\up +1.22\%})}  & {12.23{\tiny$\rightarrow$}12.35 ({\up +0.98\%})}  & {28.16{\tiny$\rightarrow$}28.44 ({\up +0.99\%})}  & {15.31{\tiny$\rightarrow$}15.44 ({\up +0.85\%})}     \\ 

~ & {SimGCL}  & {5.04{\tiny$\rightarrow$}8.49 ({\up +68.45\%})}  & {9.28{\tiny$\rightarrow$}14.86 ({\up +60.13\%})}  	& {13.56{\tiny$\rightarrow$}21.31 ({\up +57.15\%})}  & {10.33{\tiny$\rightarrow$}16.20 ({\up +56.82\%})}  & {22.70{\tiny$\rightarrow$}32.82 ({\up +44.58\%})}  & {14.31{\tiny$\rightarrow$}21.21 ({\up +48.22\%})}  & {31.68{\tiny$\rightarrow$}41.79 ({\up +31.91\%})}  & {17.54{\tiny$\rightarrow$}24.37 ({\up +38.94\%})}     \\

~ & {NCL}  & {4.10{\tiny$\rightarrow$}4.63 ({\up +12.93\%})}  & {8.21{\tiny$\rightarrow$}9.11 ({\up +10.84\%})}  	& {11.48{\tiny$\rightarrow$}13.02 ({\up +13.41\%})}  & {8.98{\tiny$\rightarrow$}10.08 ({\up +12.25\%})}  & {19.55{\tiny$\rightarrow$}22.32 ({\up +14.17\%})}  & {12.59{\tiny$\rightarrow$}14.18 ({\up +12.63\%})}  & {27.85{\tiny$\rightarrow$}31.76 ({\up +14.04\%})}  & {15.60{\tiny$\rightarrow$}17.58 ({\up +12.69\%})}     \\ 

~ & {BUIR}  & {1.16{\tiny$\rightarrow$}1.22 ({\up +5.17\%})}  & {2.55{\tiny$\rightarrow$}2.64 ({\up +3.53\%})}  	& {4.09{\tiny$\rightarrow$}4.31 ({\up +5.38\%})}  & {3.16{\tiny$\rightarrow$}3.28 ({\up +3.80\%})}  & {8.29{\tiny$\rightarrow$}8.77 ({\up +5.79\%})}  & {5.05{\tiny$\rightarrow$}5.26 ({\up +4.16\%})}  & {13.34{\tiny$\rightarrow$}14.16 ({\up +6.15\%})}  & {6.92{\tiny$\rightarrow$}7.23 ({\up +4.48\%})}     \\ 

~ & {DirectAU}  & {8.03{\tiny$\rightarrow$}9.12 ({\up +13.57\%})}  & {13.34{\tiny$\rightarrow$}14.84 ({\up +11.24\%})}  	& {21.32{\tiny$\rightarrow$}24.11 ({\up +13.09\%})}  & {15.40{\tiny$\rightarrow$}17.26 ({\up +12.08\%})}  & {34.53{\tiny$\rightarrow$}38.47 ({\up +11.47\%})}  & {20.93{\tiny$\rightarrow$}23.30 ({\up +11.32\%})}  & {46.73{\tiny$\rightarrow$}51.57 ({\up +10.36\%})}  & {25.21{\tiny$\rightarrow$}27.88 ({\up +10.59\%})}     \\ 

\midrule[0.1pt]

\multirow{7}{*}{Yelp} & {LightGCN}  & {2.13{\tiny$\rightarrow$}2.37 ({\up +11.27\%})}  & {1.31{\tiny$\rightarrow$}1.64 ({\up +25.19\%})}  	& {2.16{\tiny$\rightarrow$}2.72 ({\up +25.93\%})}  & {1.94{\tiny$\rightarrow$}2.39 ({\up +23.20\%})}  & {4.07{\tiny$\rightarrow$}5.23 ({\up +28.50\%})}  & {2.88{\tiny$\rightarrow$}3.62 ({\up +25.69\%})}  & {6.47{\tiny$\rightarrow$}8.37 ({\up +29.37\%})}  & {3.87{\tiny$\rightarrow$}4.88 ({\up +26.10\%})}     \\ 


~ & {SGL}  & {0.92{\tiny$\rightarrow$}0.97 ({\up +5.43\%})}  & {2.43{\tiny$\rightarrow$}2.50 ({\up +2.88\%})}  	& {2.28{\tiny$\rightarrow$}2.57 ({\up +12.72\%})}  & {2.14{\tiny$\rightarrow$}2.33 ({\up +8.88\%})}  & {4.06{\tiny$\rightarrow$}4.67 ({\up +15.02\%})}  & {3.02{\tiny$\rightarrow$}3.36 ({\up +11.26\%})}  & {6.21{\tiny$\rightarrow$}7.18 ({\up +15.62\%})}  & {3.91{\tiny$\rightarrow$}4.38 ({\up +12.02\%})}     \\ 

~ & {SimGCL}  & {1.01{\tiny$\rightarrow$}2.00 ({\up +98.02\%})}  & {2.72{\tiny$\rightarrow$}5.24 ({\up +92.65\%})}  	& {2.30{\tiny$\rightarrow$}5.31 ({\up +130.87\%})}  & {2.22{\tiny$\rightarrow$}4.83 ({\up +117.57\%})}  & {3.67{\tiny$\rightarrow$}8.68 ({\up +136.51\%})}  & {2.92{\tiny$\rightarrow$}6.52 ({\up +123.29\%})}  & {4.96{\tiny$\rightarrow$}11.71 ({\up +136.09\%})}  & {3.46{\tiny$\rightarrow$}7.78 ({\up +124.86\%})}     \\

~ & {NCL}  & {1.16{\tiny$\rightarrow$}1.41 ({\up +21.55\%})}  & {3.43{\tiny$\rightarrow$}4.06 ({\up +18.37\%})}  & {2.91{\tiny$\rightarrow$}3.65 ({\up +25.43\%})}  & {2.87{\tiny$\rightarrow$}3.52 ({\up +22.65\%})}  & {4.76{\tiny$\rightarrow$}6.13 ({\up +28.78\%})}  & {3.83{\tiny$\rightarrow$}4.79 ({\up +25.07\%})}  & {6.74{\tiny$\rightarrow$}8.88 ({\up +31.75\%})}  & {4.67{\tiny$\rightarrow$}5.91 ({\up +26.55\%})}   \\

~ & {BUIR}  & {0.53{\tiny$\rightarrow$}0.53 ({+0\%})}  & {1.61{\tiny$\rightarrow$}1.62 ({\up +0.62\%})}  	& {1.55{\tiny$\rightarrow$}1.58 ({\up +1.94\%})}  & {1.49{\tiny$\rightarrow$}1.51 ({\up +1.34\%})}  & {2.76{\tiny$\rightarrow$}2.84 ({\up +2.90\%})}  & {2.13{\tiny$\rightarrow$}2.17 ({\up +1.88\%})}  & {4.19{\tiny$\rightarrow$}4.33 ({\up +3.34\%})}  & {2.74{\tiny$\rightarrow$}2.79 ({\up +1.82\%})}     \\ 

~ & {DirectAU}  & {1.39{\tiny$\rightarrow$}1.66 ({\up +19.42\%})}  & {3.54{\tiny$\rightarrow$}4.12 ({\up +16.38\%})}  	& {3.36{\tiny$\rightarrow$}4.15 ({\up +23.51\%})}  & {3.11{\tiny$\rightarrow$}3.74 ({\up +20.26\%})}  & {5.66{\tiny$\rightarrow$}7.05 ({\up +24.56\%})}  & {4.24{\tiny$\rightarrow$}5.15 ({\up +21.46\%})}  & {8.28{\tiny$\rightarrow$}10.27 ({\up +24.03\%})}  & {5.31{\tiny$\rightarrow$}6.46 ({\up +21.66\%})}     \\

\midrule[0.1pt]

\multirow{7}{*}{Kindle} & {LightGCN}  & {7.21{\tiny$\rightarrow$}7.26 ({\up +0.69\%})}  & {8.81{\tiny$\rightarrow$}8.56 ({\down -2.84\%})}  	& {14.88{\tiny$\rightarrow$}15.06 ({\up +1.21\%})}  & {10.26{\tiny$\rightarrow$}10.21 ({\down -0.49\%})}  & {19.76{\tiny$\rightarrow$}20.23 ({\up +2.38\%})}  & {12.18{\tiny$\rightarrow$}12.19 ({\up +0.08\%})}  & {23.20{\tiny$\rightarrow$}24.01 ({\up +3.49\%})}  & {13.28{\tiny$\rightarrow$}13.37 ({\up +0.68\%})}     \\

~ & {SGL}  & {7.90{\tiny$\rightarrow$}7.94 ({\up +0.51\%})}  & {9.63{\tiny$\rightarrow$}9.33 ({\down -3.12\%})}  	& {16.93{\tiny$\rightarrow$}17.58 ({\up +3.84\%})}  & {11.53{\tiny$\rightarrow$}11.61 ({\up +0.69\%})}  & {23.32{\tiny$\rightarrow$}24.37 ({\up +4.50\%})}  & {13.99{\tiny$\rightarrow$}14.19 ({\up +1.43\%})}  & {27.87{\tiny$\rightarrow$}29.43 ({\up +5.60\%})}  & {15.41{\tiny$\rightarrow$}15.74 ({\up +2.14\%})}     \\ 

~ & {SimGCL}  & {8.54{\tiny$\rightarrow$}8.80 ({\up +3.04\%})}  & {11.55{\tiny$\rightarrow$}11.25 ({\down -2.60\%})}  	& {17.27{\tiny$\rightarrow$}18.57 ({\up +7.53\%})}  & {12.62{\tiny$\rightarrow$}12.98 ({\up +2.85\%})}  & {22.79{\tiny$\rightarrow$}24.71 ({\up +8.42\%})}  & {14.82{\tiny$\rightarrow$}15.38 ({\up +3.78\%})}  & {25.95{\tiny$\rightarrow$}28.20 ({\up +8.67\%})}  & {16.87{\tiny$\rightarrow$}17.50 ({\up +3.73\%})}     \\ 

~ & {NCL}   & {9.26{\tiny$\rightarrow$}9.66 ({\up +4.32\%})}  & {12.55{\tiny$\rightarrow$}12.85 ({\up +2.39\%})}  	& {18.25{\tiny$\rightarrow$}19.42 ({\up +6.41\%})}  & {13.50{\tiny$\rightarrow$}14.10 ({\up +4.44\%})}  & {23.89{\tiny$\rightarrow$}25.43 ({\up +6.45\%})}  & {15.83{\tiny$\rightarrow$}16.58 ({\up +4.74\%})}  & {27.62{\tiny$\rightarrow$}29.60 ({\up +7.17\%})}  & {17.08{\tiny$\rightarrow$}17.96 ({\up +5.15\%})}     \\ 

~ & {BUIR}  & {7.30{\tiny$\rightarrow$}7.43 ({\up +1.78\%})}  & {10.01{\tiny$\rightarrow$}10.01 ({0\%})}  	  & {15.50{\tiny$\rightarrow$}15.67 ({\up +1.10\%})}  & {11.28{\tiny$\rightarrow$}11.34 ({\up +0.53\%})}  & {20.64{\tiny$\rightarrow$}20.98 ({\up +1.65\%})}  & {13.46{\tiny$\rightarrow$}13.58 ({\up +0.89\%})}  & {24.20{\tiny$\rightarrow$}24.72 ({\up +2.15\%})} & {14.68{\tiny$\rightarrow$}14.85 ({\up +1.16\%})} \\

~ & {DirectAU}  & {8.00{\tiny$\rightarrow$}8.26 ({\up +3.25\%})}  & {10.24{\tiny$\rightarrow$}10.23 ({\down -0.10\%})}  	& {17.69{\tiny$\rightarrow$}18.66 ({\up +5.48\%})}  & {12.23{\tiny$\rightarrow$}12.60 ({\up +3.03\%})}  & {24.89{\tiny$\rightarrow$}26.57 ({\up +6.75\%})}  & {15.07{\tiny$\rightarrow$}15.67 ({\up +3.98\%})}  & {30.18{\tiny$\rightarrow$}32.61 ({\up +8.05\%})}  & {16.79{\tiny$\rightarrow$}17.60 ({\up +4.82\%})}     \\

\midrule[0.1pt]

\multirow{7}{*}{AMZ-Book} & {LightGCN}  & {2.02{\tiny$\rightarrow$}2.14 ({\up +5.94\%})}  & {4.62{\tiny$\rightarrow$}4.77 ({\up +3.25\%})}  	& {5.15{\tiny$\rightarrow$}5.65 ({\up +9.71\%})}  & {4.48{\tiny$\rightarrow$}4.82 ({\up +7.59\%})}  & {8.13{\tiny$\rightarrow$}9.26 ({\up +13.90\%})}  & {5.92{\tiny$\rightarrow$}6.49 ({\up +9.63\%})}  & {11.05{\tiny$\rightarrow$}12.78 ({\up +15.66\%})}  & {7.06{\tiny$\rightarrow$}7.85 ({\up +11.19\%})}     \\


~ & {SGL}  & {2.47{\tiny$\rightarrow$}2.59 ({\up +4.86\%})}  & {5.51{\tiny$\rightarrow$}5.62 ({\up +2.00\%})}  	& {6.06{\tiny$\rightarrow$}6.64 ({\up +9.57\%})}  & {5.30{\tiny$\rightarrow$}5.65 ({\up +6.60\%})}  & {9.46{\tiny$\rightarrow$}10.66 ({\up +12.68\%})}  & {6.94{\tiny$\rightarrow$}7.52 ({\up +8.36\%})}  & {12.64{\tiny$\rightarrow$}14.39 ({\up +13.84\%})}  & {8.18{\tiny$\rightarrow$}8.95 ({\up +9.41\%})}     \\ 

~ & {SimGCL}  & {2.72{\tiny$\rightarrow$}3.09 ({\up +13.60\%})}  & {6.40{\tiny$\rightarrow$}7.02 ({\up +9.69\%})}  	& {6.13{\tiny$\rightarrow$}7.14 ({\up +16.48\%})}  & {5.64{\tiny$\rightarrow$}6.42 ({\up +13.83\%})}  & {8.69{\tiny$\rightarrow$}10.35 ({\up +19.10\%})}  & {6.19{\tiny$\rightarrow$}7.99 ({\up +29.08\%})}  & {10.55{\tiny$\rightarrow$}12.76 ({\up +20.95\%})}  & {7.68{\tiny$\rightarrow$}8.95 ({\up +16.54\%})}     \\ 

~ & {NCL}  & {2.63{\tiny$\rightarrow$}2.97 ({\up +7.53\%})}  & {6.49{\tiny$\rightarrow$}7.18 ({\up +4.81\%})}  	& {6.30{\tiny$\rightarrow$}7.32 ({\up +8.40\%})}  & {5.81{\tiny$\rightarrow$}6.61 ({\up +6.79\%})}  & {9.53{\tiny$\rightarrow$}11.35 ({\up +9.65\%})}  & {7.44{\tiny$\rightarrow$}8.58 ({\up +7.42\%})}  & {12.44{\tiny$\rightarrow$}15.10 ({\up +10.24\%})}  & {8.58{\tiny$\rightarrow$}10.07 ({\up +7.93\%})}     \\

~ & {BUIR}  & {1.43{\tiny$\rightarrow$}1.46 ({\up +2.10\%})}  & {3.78{\tiny$\rightarrow$}3.82 ({\up +1.06\%})}  	& {3.95{\tiny$\rightarrow$}4.00 ({\up +1.27\%})}  & {3.60{\tiny$\rightarrow$}3.63 ({\up +0.83\%})}  & {6.30{\tiny$\rightarrow$}6.40 ({\up +1.59\%})}  & {4.79{\tiny$\rightarrow$}4.83 ({\up +0.84\%})}  & {8.45{\tiny$\rightarrow$}8.55 ({\up +1.18\%})}  & {5.66{\tiny$\rightarrow$}5.70 ({\up +0.71\%})}     \\ 

~ & {DirectAU}  & {2.88{\tiny$\rightarrow$}3.13 ({\up +8.68\%})}  & {6.55{\tiny$\rightarrow$}7.00 ({\up +6.87\%})}  	& {6.88{\tiny$\rightarrow$}7.63 ({\up +10.90\%})}  & {6.12{\tiny$\rightarrow$}6.68 ({\up +9.15\%})}  & {10.48{\tiny$\rightarrow$}11.80 ({\up +12.60\%})}  & {7.89{\tiny$\rightarrow$}8.70 ({\up +10.27\%})}  & {13.68{\tiny$\rightarrow$}15.54 ({\up +13.60\%})}  & {9.16{\tiny$\rightarrow$}10.17 ({\up +11.03\%})}     \\

\bottomrule
\end{tabular}}
\vspace{-0.1in}
\end{table*}

From Table~\ref{tab:exp}, we notice that
\begin{itemize}[leftmargin=*]
\item After integrating \model, recent recommender models improve their segmentation capability across all datasets.
Not only does this show our method's effectiveness, but more importantly, this also validates its generality and flexibility to the variety of graph-based models as well as different datasets.

\item We notice that the model improvements on MovieLens dataset are larger than those on other datasets.
One major explanation is that users of MovieLens have more average interactions, i.e., 165.60 as shown in Table~\ref{tab:datasets}, leading to more complicated user preference distributions whereas our \model~can well utilize such rich information to encode the user-wise similarity.

\item Furthermore, equipped with \model, contrastive-learning-based models, e.g., SGL~\cite{sgl}, SimGCL~\cite{simgcl}, and NCL~\cite{ncl}, generally have larger model improvements.
This is because augmentation techniques (either to original data or to the latent features) subsequently provide the embedding enrichment for \model~ to exert.
\end{itemize}

\begin{figure}[t]
\centering
\begin{minipage}{0.5\textwidth}
\includegraphics[width=3.4in]{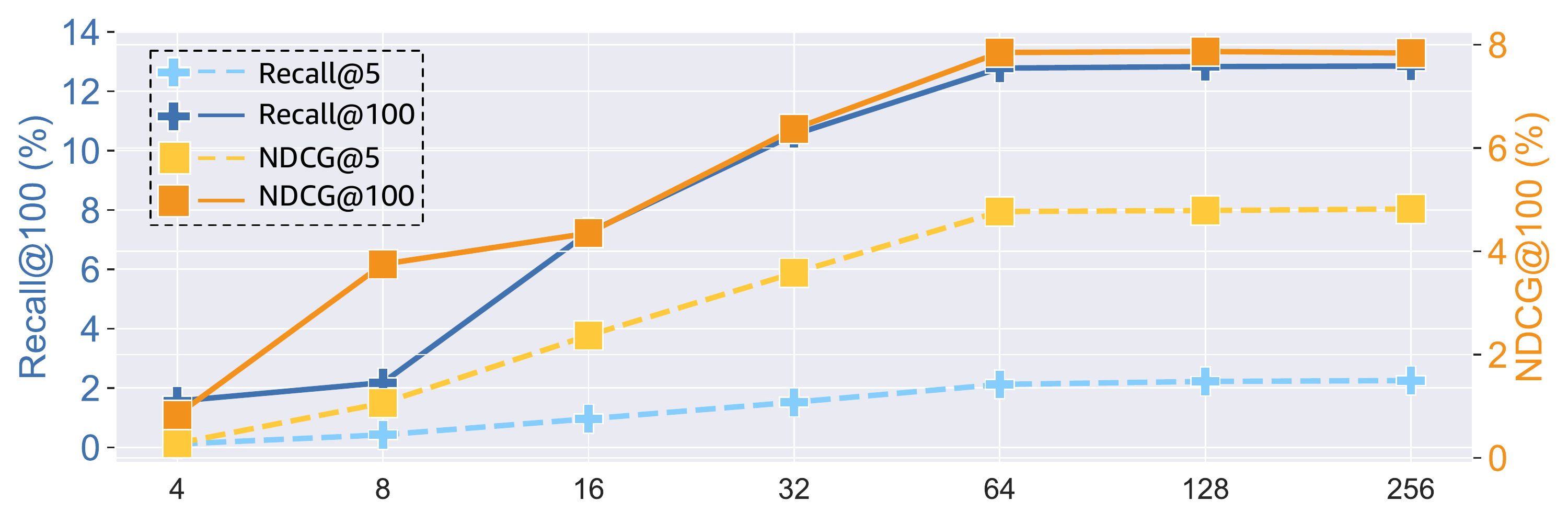}
\end{minipage} 
\vspace{-0.2in}
\caption{Experimental results of altering $S$ values.}
\label{fig:ns}
\end{figure}

\subsubsection{\textbf{Effect of Slicing Number $S$ (RQ2A).}}
\label{sec:change_S}
Due to the renowned and stable performance of LightGCN~\cite{lightgcn}, we utilize it as the backbone on AMZ-Book dataset to exemplify the model analysis.
We alternatively change the value of $S$ and plot the results in Figure~\ref{fig:ns}.
We notice that, altering $S$ from $4$ to $64$ is more influential to the model performance, which is intuitive as this produces a more accurate and fine-grained cumulative approximation.
However, on the other hand, consistently increasing $S$ will also put more computation and memory strains.
Thus, setting $S$ as 64 is the balanced spot with positive momentum that presents a practical trade-off between model performance and resource consumption.

\subsubsection{\textbf{Dimension Reduction (RQ2B).}}
During evaluation, we notice that some models encounter the ``out-of-memory'' problem.
To address this issue, we approach to aggregate layer-wise embeddings in Eqn.(\ref{eq:layerwise}) to reduce the total dimensionality from $S(L+1)$ to $S$.
\begin{wraptable}[6]{l}{0.27\textwidth}
\centering
\footnotesize
\caption{Aggregation Selection.}
\vspace{-0.15in}
\label{tab:agg}
\setlength{\tabcolsep}{1mm}{
\begin{tabular}{c | c  c | c  c| c  c }
\toprule 
             Aggregator     &  \multicolumn{2}{c|}{\footnotesize \textit{Concat}} & \multicolumn{2}{c|}{\footnotesize \textit{Sum}}   &  \multicolumn{2}{c}{\footnotesize \textit{Max}}  \\
             Metric & {\scriptsize K=5} & {\scriptsize K=100} & {\scriptsize K=5} & {\scriptsize K=100} &{\scriptsize K=5} & {\scriptsize K=100}  \\
\midrule[0.1pt]
  {\scriptsize Recall@K}  	& \textbf{2.14}   & {12.78}     & {2.12}   & \textbf{12.81}   & {1.77}   & {11.27}     \\ 
    {\scriptsize NDCG@K}  	& {4.77}   & \textbf{7.85}    & \textbf{4.78}   & {7.77}   & {4.31}   & {7.01}    \\
\bottomrule
\end{tabular}}
\end{wraptable}
From Table~\ref{tab:agg}, we notice that \textit{Sum} surprisingly presents a competitive performance with the original \textit{Concat} operation. 
This indicates that, while \textit{Concat} has a more complete representation encoding with theoretical supports, \textit{Sum} is suitable for dimension reduction in scenarios with limited computational resources.


\section{Conclusion and Future Extension}
In this work, we propose \model~to encode representations for effective user segmentation in collaborative filtering.
The extensive experiments demonstrate the effectiveness of our proposed method and its generality to a variety of model deployments. 
As for future work, we point out three major directions as follows:
\begin{enumerate}[leftmargin=*]
\item In light of the empirical findings in~\cref{sec:overall}, it is interesting to explore \textit{contrastive learning} techniques~\cite{yifei23,zhang2022costa,Ma2022GraphCC,liu2022hierarchical} in the sub-graph feature domain for further model improvement.

\item It is worth investigating adapting our training-free model to other scenarios of information retreival and autonomous database management~\cite{qiu2022efficient,hu2020selfore,hu2021semi,liu2022semantic,liu2023comprehensive,hu2021gradient,xin2023CAMdeep}.

\item We plan to design unsupervised regularization mechanisms such that \model~and the backbone model can be \textit{jointly optimized} or even \textit{mutually enhanced} for multi-task learning.
\end{enumerate}

\begin{acks}
Yankai Chen, Yifei Zhang, Menglin Yang, Zixing Song and Irwin King were partially supported by the National Key Research and Development Program of China (No. 2018AAA0100204) and by the Research Grants Council of the Hong Kong Special Administrative Region, China (RGC GRF 2151185; CUHK 14222922).
Chen Ma was supported by the Start-up Grant (No. 9610564) and the Strategic Research Grant (No. 7005847) of City University of Hong Kong.
\end{acks}
\balance
\bibliographystyle{ACM-Reference-Format}
\bibliography{reference}

\end{document}